\documentclass{edbk} 
  \usepackage{edbkps}

\let\footnoterule\savefootnoterule 

\kluwerbib

\input epsf

\begin{document}



\articletitle[]{Comptonization, reflection and noise in black hole
binaries\footnote{To appear in the Proceedings of the 5th CAS/MPG
Workshop on High Energy Astrophysics}}


\author{M.Gilfanov$^{1,2}$, E.Churazov$^{1,2}$, M.Revnivtsev$^{2,1}$}
\affil{$^1$Max-Planck-Institute f\"ur Astrophysik,
Karl-Schwarzschild-Str. 1, D-85740 Garching, Germany}
\affil{$^2$Space Research Institute, Profsoyuznaya 84/32, 117810 Moscow,
Russia}



\begin{abstract}

Based on the analysis of a large sample of RXTE/PCA observations of
several black hole binaries in the low spectral state we show that a
correlation exists between the spectral parameters and characteristic
noise frequency. In particular, the amplitude of reflection increases
and the slope of Comptonized radiation steepens as the noise frequency
increases. We consider possible implications of these results on the
accretion flow models and discuss a possible observational test aimed
to discriminate between different geometries of the accretion flow.

\end{abstract}



Comptonization of soft seed photons in a hot, optically thin
electron cloud near the compact object is thought to produce the hard 
X--ray radiation in the low spectral state of accreting black holes
(\cite{str79}, \cite{st80}). The slope of the Comptonized
spectrum  is governed by the ratio of the energy deposited into the
electrons and the influx of the soft radiation into the Comptonization
region; the lower the ratio the steeper the Comptonized spectrum 
(e.g. \cite{st89}, \cite{feedback}, \cite{kemer}).

\begin{figure}
\begin{centering}
\centerline{
\epsfxsize 8 cm
\epsffile{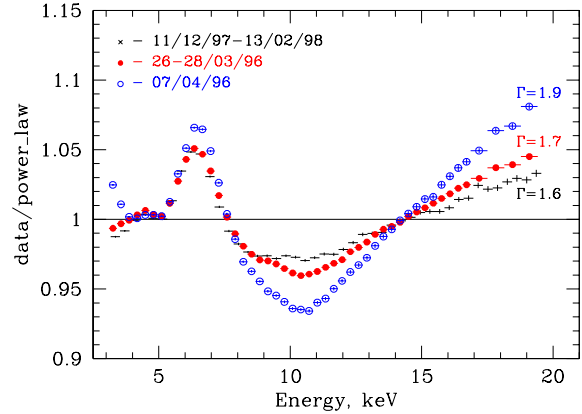}}
\end{centering}
\caption{ The counts spectra of Cyg X--1 at different epoch 
plotted as a ratio to a power law model. The power law photon index is
indicated against each spectrum.
\label{plrat}
}
\end{figure}

Reflection of the Comptonized radiation from neutral or partially
ionized matter, presumably the optically thick accretion disk, leads
to appearance of characteristic features in the spectra of X--ray
binaries (Fig.\ref{plrat}). The main signatures of the emission
reflected from cold neutral medium are well known -- the fluorescent
K$_{\alpha}$ line of iron at 6.4 keV, iron K-edge at 7.1 keV  and a
broad hump above $\sim 20-30$ keV (Basko, Sunyaev \& Titarchuk 1974,
George \& Fabian 1991). 
The exact shape of these spectral features in the X-ray binaries depends on
the ionization state of the reflecting medium (\cite{ionrefl}) and might
be modified by the strong gravity effects and intrinsic motions in the
reflector (e.g. Fabian et al., 1989). The amplitude of the reflection
signatures depends primarily on the ionization state and on the solid
angle of the reflector as seen from the source of the primary
radiation.

Based on the analysis of a large sample of Seyfert AGNs and several
X--ray binaries \cite{zdz1} (see also \cite{zdzrev}) found a
correlation between the amplitude of reflection and the slope of the
underlying power law.   They concluded that the existence of such a
correlation implies a dominant role of the reflecting medium as the
source of seed photons to the Comptonization region.

\begin{figure*}
\hbox{	
\epsfxsize 6.1 cm
\epsffile{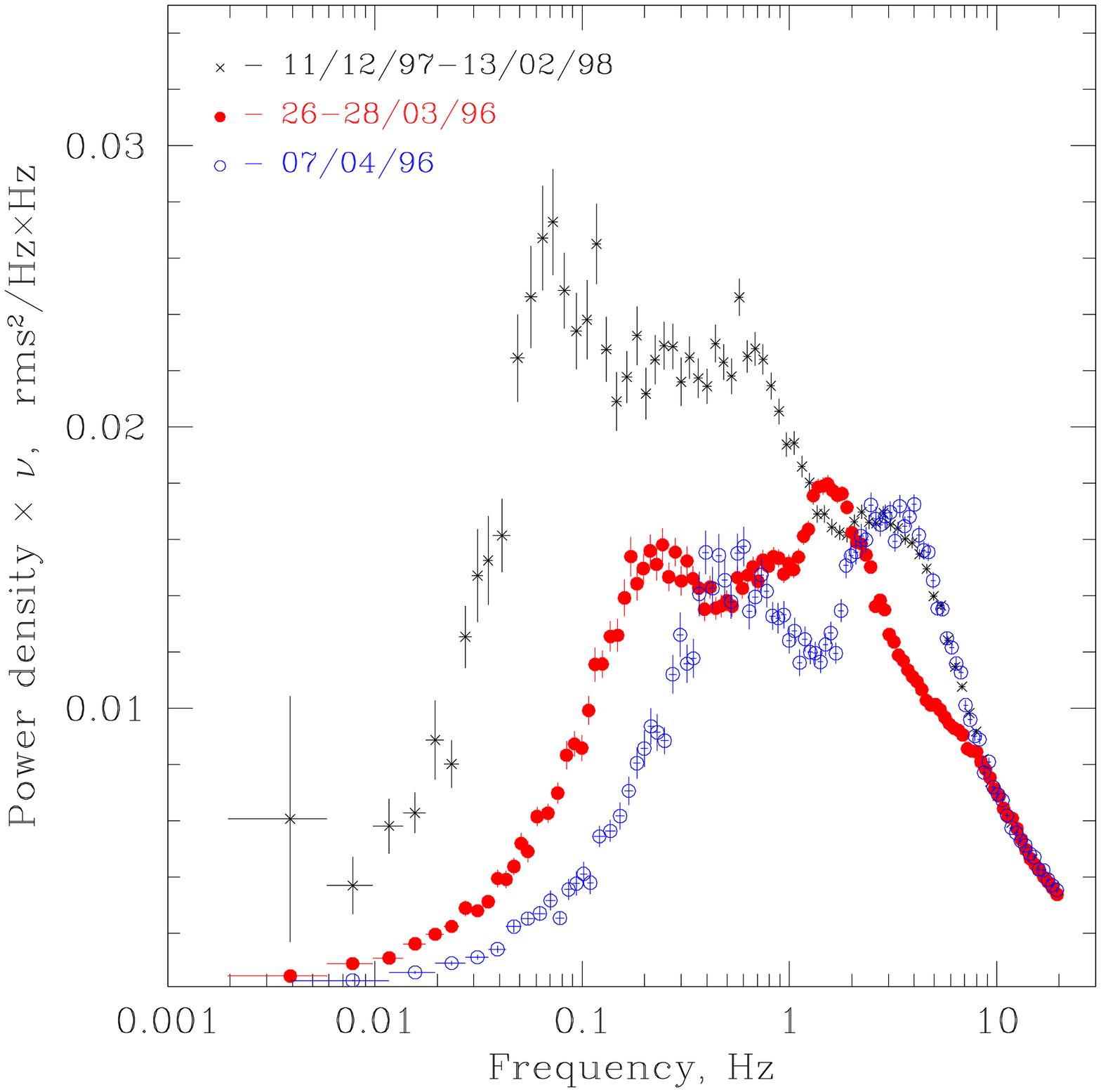}
\epsfxsize 6.1 cm
\epsffile{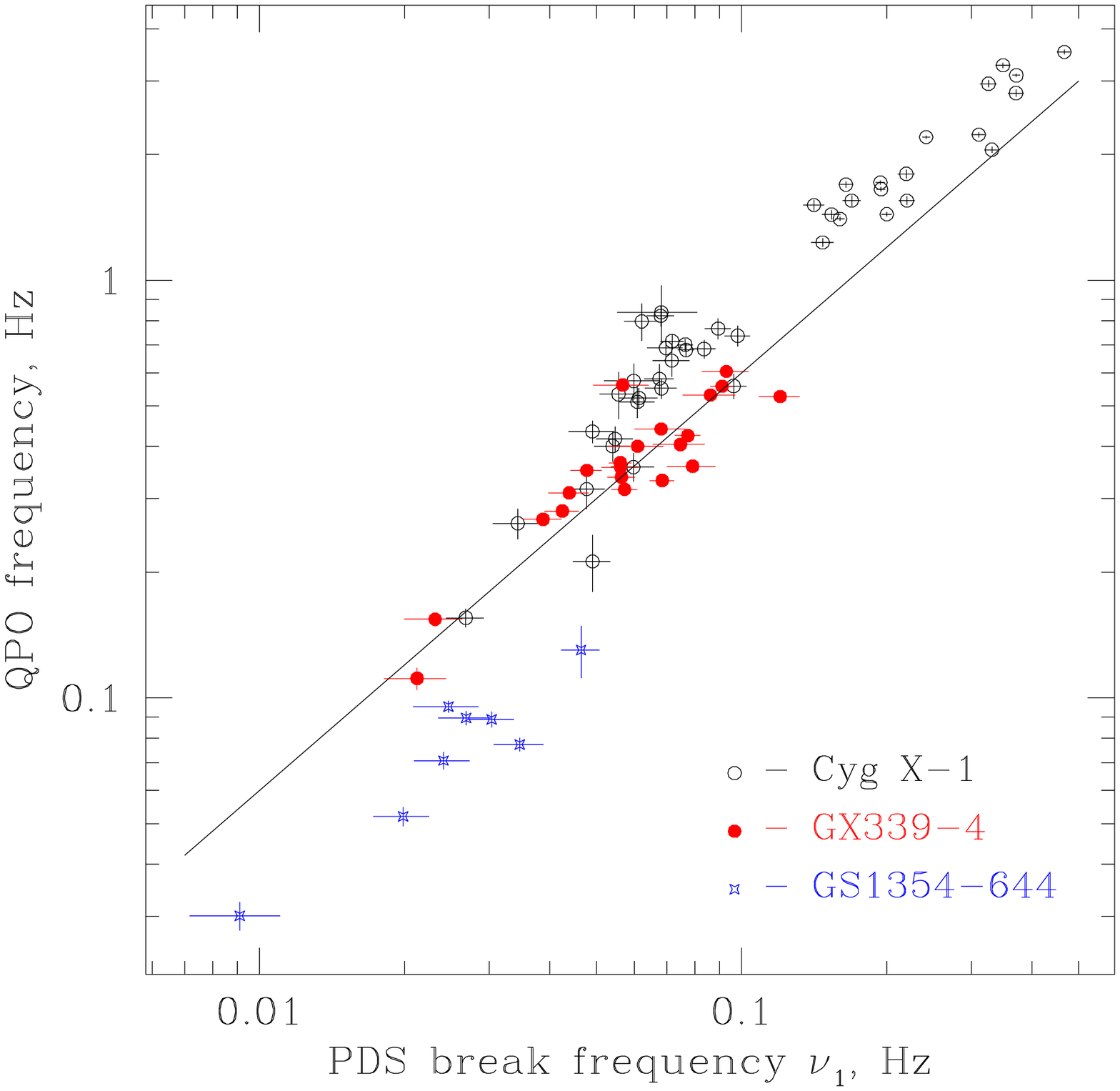}
}
\caption{ {\em Left:} The power density spectra of Cyg X--1 at
different epoch.  The power density spectra are plotted as frequency
$\times$ (power density), i.e. in units of $\rm Hz\times rms^2/Hz$.
The observations and the symbols are the same as in Fig.\ref{plrat}
{\em Right:} The relation between the QPO centroid frequency and the break
frequency of the band limited noise for Cyg X-1, GX339--4 and
GS1354--644. The solid line is $f_{qpo}=6\times f_{br}$.
\label{pds}
}
\end{figure*}

The power density spectra of the black hole binaries in the low state
(see \cite{qporev} for a review) plotted in the units of
frequency$\times$power usually appear as a superposition of two or
more rather broad humps containing most of the power of aperiodic
variations below several tens Hz (Fig.\ref{pds}).  These humps define
several characteristic noise frequencies of which the lowest usually
corresponds to the break  frequency $\nu_{br}$ of the band limited
noise component. Although a number of theoretical models was proposed
to explain the power spectra of X--ray binaries (e.g. \cite{alpar},
\cite{stella}), the nature of the characteristic noise
frequencies is still unclear.   Despite the fact that the
characteristic noise frequencies vary from source to source and from
epoch to epoch,  they are correlated with  the break frequency
of the band limited noise (Fig.\ref{pds}, \cite{br_qpo}).

\cite{paper1} showed that a tight correlation exists between the
characteristic noise frequencies  and the spectral parameters. Based
on RXTE/PCA observations of Cyg X--1 from 1996--1998 they found that
the amplitude of reflection increases and the spectrum of primary
radiation steepens as the noise frequency increases.
An identical behavior was found for another black hole candidate,
GX339-4, by \cite{gx339}.  
Interpreting these correlations \cite{paper1} suggested that increase
of the noise frequency might be caused by the shift of the inner
boundary of the  optically thick accretion disk towards the compact
object. The related increase of the solid angle, subtended by the
disk, and of the influx of the soft photons to the Comptonization
region lead to an increase of the amount of reflection and
steepening of the Comptonized spectrum.

Below we investigate further the relation between the spectral
parameters and characteristic noise frequencies based on a larger
sample of objects and observations and discuss implications of these
results on the geometry of the accretion flow in the low spectral
state of black hole binaries.

\section{The data}

\subsubsection{Observations and data reduction}

The results presented below are based on the publicly available data
of Cyg X--1, GX339--4 and GS1354--644 observations with the
Proportional Counter Array aboard the Rossi X-ray Timing Explorer
(\cite{rxte}) from 1996--1998 during the
low (hard) spectral state of the sources. In total our sample
included $\approx 60$ observations of these sources. The energy and
power density spectra were averaged for each individual observation.  
The energy spectra were extracted from the ``Standard Mode 2'' data and
ARF and RMF were constructed using standard RXTE FTOOLS v.4.2 tasks.
The ``VLE'' and ``Q6'' models were used for the background calculation
with the preference being given to the ``VLE'' model when possible.
A uniform systematic error of 0.5\% was added quadratically to the 
statistical error in each energy channel.  
The power spectra were calculated in the 2--16 keV energy
band and in the $\approx 0.005-32$ Hz frequency range  following the
standard X--ray timing technique and normalized to units of squared
fractional rms.

\subsubsection{The spectral model}

The energy spectra were fit with a model consisting of a power 
law with a superimposed reflected continuum (pexrav model in XSPEC,
\cite{pexrav}) and an intrinsically narrow line at $\sim 6.4-6.7$
keV. The centroid energy of the line  was a free parameter of the
fit. For all sources the binary system inclination was fixed at $i=50$
degrees, the iron abundance -- at the solar value of $A_{\rm
Fe}=3.3\cdot 10^{-5}$ and the low energy  absorption --  at $N_{\rm
H}=6\cdot 10^{21}$ cm$^{-2}$.  The effects of ionization were not included. 
In order to approximately include in the model the smearing of the 
reflection features due to motion in the accretion flow 
the reflection continuum and line were convolved with a Gaussian,
which width was a free parameter of the fit. 
The spectra were fitted in the 4--20 keV energy range. 
This spectral model is identical to that applied by \cite{paper1}
to  the Cyg X-1 data, except that in the present analysis we
let the line centroid energy be a free parameter of the fit. Strength
of the reflected component is characterized in the pexrav model by the
reflection scaling factor $R$, which
approximately measures the solid angle subtended by the reflecting
media as seen from the source of the primary radiation, $R\approx
\Omega_{refl}/2\pi$, so that $R=1$ for an isotropic point source above
an infinite optically thick slab.

The spectral model is obviously oversimplified. Moreover, several
important parameters, such as the binary system inclination angle and
the iron abundance were fixed at fiducial values.
Therefore the best fit values  do not necessarily represent the exact
values of the physically interesting parameters.  Particularly subject
to the uncertainties due to the choice of the spectral model is the
reflection scaling factor $R\sim\Omega/2\pi$. 
This might explain the values of $R$ exceeding unity
obtained for some of the spectra. 
However, as was shown by \cite{paper1} the model correctly ranks the
spectra according to the strength of the reflected component and the
slope of the underlying power law.

\begin{figure}[t]
\centerline{
\epsfxsize 9. cm
\epsffile{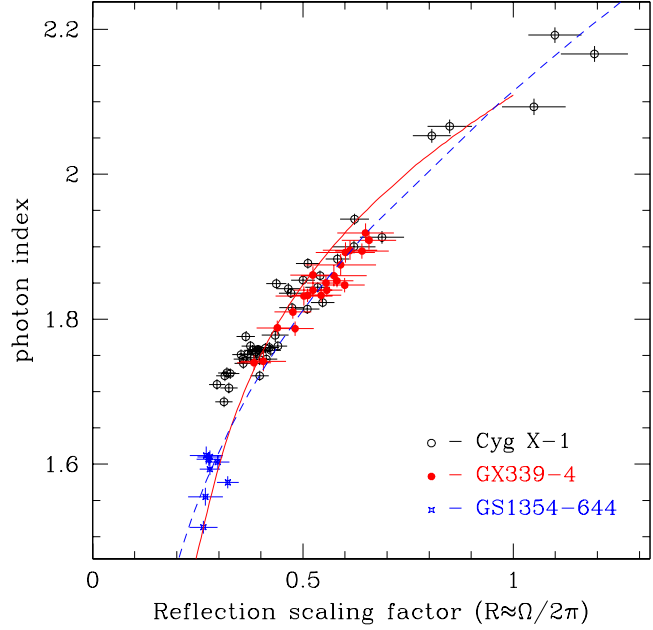}
}
\caption{The best fit values of the photon index of the underlying
power law plotted vs. reflection scaling factor for Cyg X-1, GX339--4
and GS1354--644. 
The solid line shows the dependence $\Gamma(R)$ of the photon index of
the Comptonized radiation $\Gamma$ on the strength of the reflected
emission $R$ expected in the disk--spheroid model assuming the disk
albedo $a=0.1$, Thomson optical depth of the cloud $\tau_T=1$ and the
ratio of the temperature of the seed photons to the electron
temperature $T_{bb}/T_e=10^{-4}$. The dashed line shows $\Gamma(R)$
dependence expected in the plasma ejection model (\cite{belob}) for
$a=0.15$, $\tau_T=1$, $T_{bb}/T_e=3\cdot 10^{-3}$ and $\mu_s=0.3$.
\label{refl_slope}
}
\end{figure}

\begin{figure}[t]
\centerline{	
\epsfxsize 9.0 cm
\epsffile{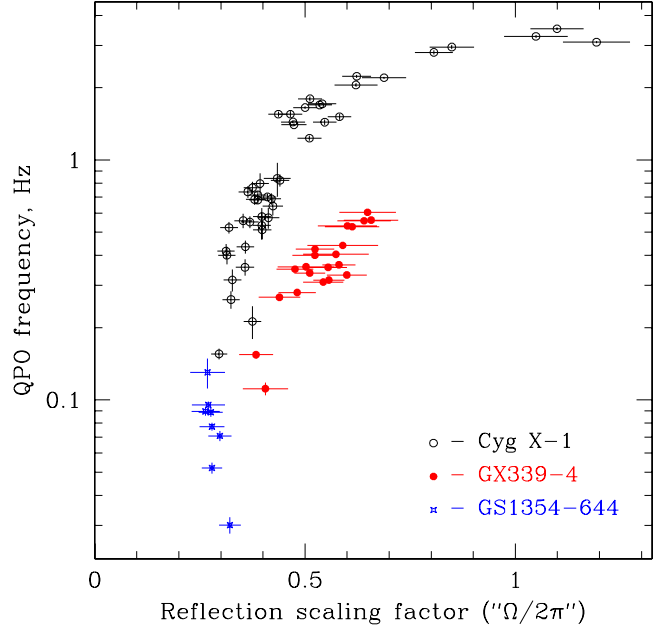}
}
\caption{The QPO centroid frequency  plotted vs. reflection scaling
factor for the same set of observations of Cyg X-1, GX339--4 and
GS1354--644 as in Fig.\ref{refl_slope} and \ref{refl_gsmo}. 
\label{refl_qpo}
}
\end{figure}

\subsubsection{Power spectra approximation}

The power density spectra in units of $rms^2/Hz$  were fit in the
0.005--15 Hz frequency range with a model, consisting of two broken
constant components and a Lorentzian profile
$P_\nu=\frac{P_0}{1+(2(\nu-\nu_0)/\Delta\nu)^2}$. We use centroid
frequency of the  Lorentzian profile  as a characteristic noise
frequency. Its dependence on the break frequency of the band limited
noise (the first broken constant component in the model) is shown in
Fig.\ref{pds}.

\subsubsection{Correlation between the spectral parameters and the
noise frequency}
The results of the energy and power spectra approximation are
presented in Fig. \ref{refl_slope} and \ref{refl_qpo}
showing dependence of the slope of the primary emission and
characteristic noise frequency on the amplitude of the reflected
component. The noise frequency and the spectral parameters are clearly
correlated. An increase of the amplitude of the reflected component is
accompanied with a steepening of the spectrum of the primary emission
and an increase of the noise frequency.

\section{Discussion}

\subsubsection{Slope of the Comptonized spectrum and reflection}

The photon index $\Gamma$ of the Comptonized radiation is governed by
the Compton amplification factor $A$ which equals to 
the ratio of the energy deposited into the hot electrons to the
influx of the soft seed photons to the Comptonization
region. The concrete shape of the $\Gamma(A)$  relation depends on the
ratio $T_{bb}/T_e$ of the temperatures of the seed photons and the
electrons, the Thomson optical depth and the geometry. In the
simplest although not unique scenario  the correlation 
between the amplitude of the reflected component $R$ and the slope of the
primary radiation $\Gamma$ (Fig.\ref{refl_slope}) could be understood
assuming that there is a positive correlation between the solid angle
subtended by the reflecting media, $\Omega_{refl}$, and the influx of
the soft photons to the Comptonization region (\cite{zdz1}). 
Existence of such correlation is a strong argument in
favor the reflecting media being the primary source of the soft seed
photons to the Comptonization region. In the absence of strong beaming
effects a correlation between $\Omega_{refl}$ and the seed photons
flux should be expected since an increase of the solid angle of the
disk seen by the hot electrons ($=\Omega_{refl}$) should generally
lead to the increase of the fraction of the disk emission reaching the 
Comptonization region.

\begin{figure}[t]
\centerline{
\epsfxsize 9. cm
\epsffile{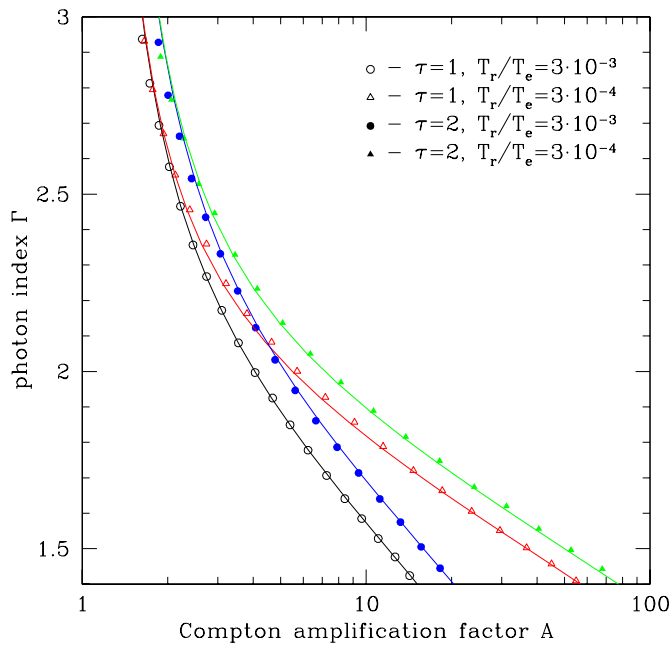}
}
\caption{Relation between the Compton amplification factor and photon
index of Comptonized radiation for an isotropic point source of the
soft seed photons with temperature $T_r$ in the center of a uniform
sphere with electron temperature $T_e$ and Thomson optical depth
$\tau_T=1$ (hollow symbols) and $\tau_T=2$ (solid symbols). The
symbols are a result of the Monte-Carlo calculations, the solid curves 
were calculated according to the formula given in the text.
\label{enh_slope}
}
\end{figure}

In order to demonstrate that such geometrical effects can
explain the observed dependence $\Gamma(R)$ we consider two idealized 
models having different cause of change of the $\Omega_{refl}$.  
In the first, disk-spheroid model, an optically thin
uniform hot sphere with radius $R_{sph}$ is surrounded with an
optically thick cold disk with an inner radius $R_{disk}$
(e.g. \cite{sombrero}), the $\Omega_{refl}$ depending on the ratio
$R_{disk}/R_{sph}$. Propagation of the disk towards/inwards the hot  
sphere (decrease of $R_{disk}/R_{sph}$) leads to an increase of the
reflection scaling factor, a decrease of the the Compton amplification
factor $A$ and a steepening of the Comptonized spectrum.
In such context the model was first studied by \cite{zdz1} and we
used their results to calculate the relation between the reflection
scaling factor $R$ and the Compton amplification $A$. In the second,
plasma ejection model, value of the $\Omega_{refl}$ is defined by the
intrinsic properties of the emitting hot plasma, particularly by its
bulk motion with mildly relativistic velocity towards or away from the
disk, which itself remains unchanged (\cite{belob}).  In the case of
an infinite disk, values of reflection below and above unity correspond
to the hot plasma moving respectively away and towards the disk.
Both models predict relation between
reflection $R$ and Compton amplification factor $A$ which can be
translated to $\Gamma(R)$  given a dependence $\Gamma(A)$ of the
photon index of Comptonized spectrum on the amplification factor.  
The latter was approximated by:
$$
A=(1-e^{-\tau_T})\cdot\frac{1-\Gamma}{2-\Gamma}\cdot
\frac{\left(\frac{T_e}{T_{bb}}\right)^{2-\Gamma}-1}{\left(\frac{T_e}{T_{bb}}\right)^{1-\Gamma}-1}+e^{-\tau_T}
$$
This formula is based on  a representation of the Comptonized spectrum
by a power law in the energy range $3kT_{bb}-3kT_e$ and agrees with the
results of the Monte-Carlo calculations with reasonable accuracy for
optical depth $\tau_T\sim 1$ and $T_{bb}/T_e\sim 10^{-5}-10^{-3}$
(Fig.\ref{enh_slope}).

\begin{figure}[t]
\centerline{
\epsfxsize 9. cm
\epsffile{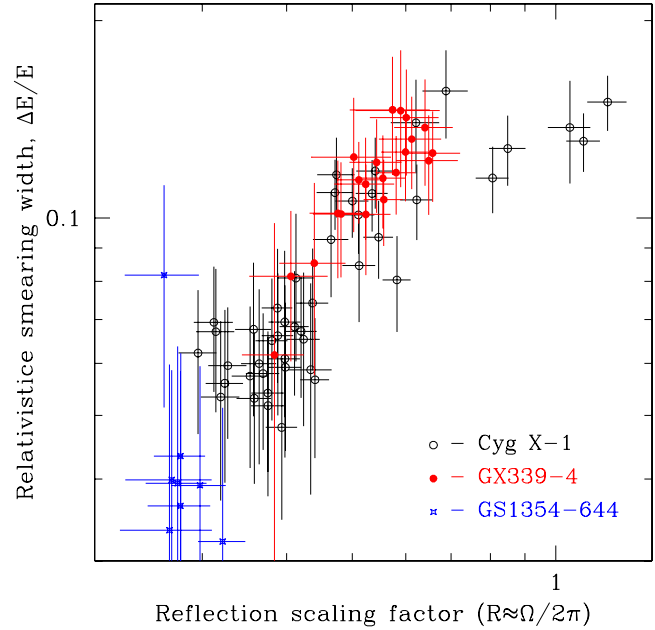}
}
\caption{The best fit value of the relative width of the relativistic
smearing of the reflection features plotted vs. reflection scaling
factor for Cyg X-1, GX339--4 and GS1354--644. 
\label{refl_gsmo}
}
\end{figure}

The expected $\Gamma(R)$ relations are shown in 
Fig.\ref{refl_slope}. With a proper tuning of the parameters both
models can reproduce the observed shape of the $\Gamma(R)$ dependence
and in this respect are virtually indistinguishable. 
The observed range of the reflection $R\sim 0.3-1$ and the slope
$\Gamma\sim 1.5-2.2$ can be explained assuming variation of the disk
radius in from $R_{disk}\sim R_{sph}$ to $R_{disk}\sim 0$ in the
disk-spheroid model or variation of the bulk motion velocity from
$v\sim 0.4 c$ away from the disk to $v\sim 0$ in the plasma ejection model.  
Finally, it should be emphasized that these
idealized models do not include a number of important effects which
might affect the particular shape of the $\Gamma(R)$ dependence.

\subsubsection{Relativistic smearing and reflection}

The spectrum of emission, reflected from a Keplerian
accretion disk is expected to be modified by special and general
relativity effects (e.g. Fabian et al., 1989). If increase of the
reflection is caused by a shift of the inner boundary of the accretion
disk closer to the compact object, a correlation between the
reflection and relativistic smearing should be expected. Such a
correlation is a generic prediction of the model and might be used to
discriminate between different assumptions about geometry of the
accretion flow.   The energy resolution of RXTE/PCA is not adequate to
study relativistic smearing of the reflection features with sufficient 
degree of confidence. However, the data shown in Fig.\ref{refl_gsmo}
might present an evidence in favor of correlated behavior of the
reflection and relativistic smearing. If confirmed, it would prove
that the spectral and timing properties in the low state of the black
hole binaries  are defined mainly by the position of the inner
boundary of the optically thick accretion disk.

\section{Conclusions}

In the low spectral state of black hole candidates the
amplitude of reflection $R$, slope of the Comptonized radiation
$\Gamma$ and characteristic noise frequency change in a correlated
way. The correlation between the slope and reflection indicates that
the accretion disk is the primary source of the soft seed photons for
Comptonization. We considered two idealized models, the disk--spheroid
and plasma ejection model, and showed that with appropriate tuning of
the parameters either model can explain the observed  $\Gamma(R)$ dependence.
The RXTE data also hints at a correlated change of the
width of relativistic smearing of the reflection features. If
confirmed by observations with better energy resolution, this
correlation may help to distinguish between different geometries of
the accretion flow.



\begin{acknowledgments}

This research has made use of data obtained through the High Energy
Astrophysics Science Archive Research Center Online Service, provided 
by the NASA/Goddard Space Flight Center.  
The work was done in the context of the research network
"Accretion onto black holes, compact objects and protostars"
(TMR Grant ERB-FMRX-CT98-0195 of the European Commission).
M.Revnivtsev acknowledges
hospitality of the Max--Planck Institute for Astrophysics and a partial
support by RBRF grant 96--15--96343 and INTAS grant 93--3364--ext.

\end{acknowledgments}



%





\begin{chapthebibliography}{}


\bibitem[Alpar et al. 1992]{alpar} 
Alpar M. A., Hasinger G., Shaham J.,  Yancopoulos S., 1992, A\&A
257,  627 

\bibitem[Basko, Sunyaev \& Titarchuk 1974]{bst} Basko M., Sunyaev
R.,  Titarchuk L., 1974, A\&A 31, 249

\bibitem[Beloborodov 1999]{belob} Beloborodov, A. M. 1999, 
ApJ, 510, L123 

\bibitem[Brandt Rotschild \& Swank 1996]{rxte} Brandt, H., 
Rotschild, R. \& Swank, J. 1996, Memorie della Societa Astronomica 
Italiana, 67, 593 

\bibitem[George \& Fabian 1991]{fab} George I.M., Fabian A.C., 1991,
MNRAS 249, 352

\bibitem[Gilfanov et al. 1995]{kemer} Gilfanov M., Churazov E.,
Sunyaev R. et al.,  1995, in: The Lives of the Neutron Stars. 
Proceedings of the NATO ASI, eds. M.A. Alpar,
U. Kiziloglu, J. van  Paradijs; Kluwer Academic,  p.331

\bibitem[Gilfanov, Churazov \& Revnivtsev 1999]{paper1} Gilfanov M.,
Churazov E., Revnivtsev M.,  1999 A\&A, 352, 182

\bibitem[Haardt \& Maraschi 1993]{feedback} Haardt F., 
Maraschi L.,  1993, ApJ 413, 507 

\bibitem[Magdziarz \& Zdziarski 1995]{pexrav} Magdziarz, P.  
\& Zdziarski, A. A. 1995, MNRAS, 273, 837 

\bibitem[Poutanen Krolik \& Ryde 1997]{sombrero} Poutanen, J. , 
Krolik, J. H. \& Ryde, F.  1997, MNRAS, 292, L21 

\bibitem[Revnivtsev, Gilfanov \& Churazov 2000]{gx339} Revnivtsev
M., Gilfanov M., Churazov E., 2000, A\&A Letters, submitted

\bibitem[Ross, Fabian \& Young 1999]{ionrefl} Ross, R. R., 
Fabian, A. C. \& Young, A. J. 1999, MNRAS, 306, 461 

\bibitem[Stella Vietri \& Morsink 1999]{stella} Stella, L. , 
Vietri, M.  \& Morsink, S. M. 1999, ApJ, 524, L63 

\bibitem[Sunyaev \& Truemper 1979]{str79} Sunyaev R., Truemper J.
1979, Nature 279, 506 

\bibitem[Sunyaev \& Titarchuk 1980]{st80} Sunyaev R., Titarchuk L.,
1980,  A\&A 86, 121 

\bibitem[Sunyaev \& Titarchuk 1989]{st89}
Sunyaev R.,  Titarchuk L., 1989,  in Proceedings of  ``23rd ESLAB
Symposium'', ESA SP-296, Bologna, Italy, eds. J.Hunt, B.Battrick,
v.1, p.627 

\bibitem[van der Klis 1995]{qporev}van der Klis M., 1995, in: X--ray
Binaries, eds. Lewin W., van Paradijs J., van den Heuvel E.P.J.;
Cambridge Univ.Press, p.252

\bibitem[Wijnands \& van der Klis 1999]{br_qpo} Wijnands R.,  
van der Klis M.,  1999, ApJ 514, 939 

\bibitem[Zdziarski, Lubinski \& Smith 1999]{zdz1} Zdziarski A.A.,
Lubinski P., Smith D.,  1999, MNRAS 303, L11

\bibitem[Zdziarski 2000]{zdzrev} Zdziarski A.A., in: Highly Energetic
Physical Processes and Mechanisms for Emission from Astrophysical
Plasmas. IAU Symp., Vol. 195, 2000, Eds: P.C.H.Martens \& S.Tsuruta


\end{chapthebibliography}
\end{document}